\documentclass[aps,prl,superscriptaddress,showpacs,floatfix,twocolumn]{revtex4}

\usepackage{graphicx}   

\begin{document}

\title{Energy dependence of $\pi^0$ production in Cu+Cu collisions \\
	at $\sqrt{s_\mathrm{NN}}$ = 22.4, 62.4, and 200~GeV}

\newcommand{\abilene}{Abilene Christian University, Abilene, TX 79699, U.S.}
\newcommand{\acadsin}{Institute of Physics, Academia Sinica, Taipei 11529, Taiwan}
\newcommand{\banaras}{Department of Physics, Banaras Hindu University, Varanasi 221005, India}
\newcommand{\barc}{Bhabha Atomic Research Centre, Bombay 400 085, India}
\newcommand{\bnlchem}{Chemistry Department, Brookhaven National Laboratory, Upton, NY 11973-5000, U.S.}
\newcommand{\bnlcoll}{Collider-Accelerator Department, Brookhaven National Laboratory, Upton, NY 11973-5000, U.S.}
\newcommand{\bnlphys}{Physics Department, Brookhaven National Laboratory, Upton, NY 11973-5000, U.S.}
\newcommand{\caucr}{University of California - Riverside, Riverside, CA 92521, U.S.}
\newcommand{\charlesczech}{Charles University, Ovocn\'{y} trh 5, Praha 1, 116 36, Prague, Czech Republic}
\newcommand{\ciae}{China Institute of Atomic Energy (CIAE), Beijing, People's Republic of China}
\newcommand{\cns}{Center for Nuclear Study, Graduate School of Science, University of Tokyo, 7-3-1 Hongo, Bunkyo, Tokyo 113-0033, Japan}
\newcommand{\colorado}{University of Colorado, Boulder, CO 80309, U.S.}
\newcommand{\columbia}{Columbia University, New York, NY 10027 and Nevis Laboratories, Irvington, NY 10533, U.S.}
\newcommand{\czechtech}{Czech Technical University, Zikova 4, 166 36 Prague 6, Czech Republic}
\newcommand{\dapnia}{Dapnia, CEA Saclay, F-91191, Gif-sur-Yvette, France}
\newcommand{\debrecen}{Debrecen University, H-4010 Debrecen, Egyetem t{\'e}r 1, Hungary}
\newcommand{\elte}{ELTE, E{\"o}tv{\"o}s Lor{\'a}nd University, H - 1117 Budapest, P{\'a}zm{\'a}ny P. s. 1/A, Hungary}
\newcommand{\fit}{Florida Institute of Technology, Melbourne, FL 32901, U.S.}
\newcommand{\fsu}{Florida State University, Tallahassee, FL 32306, U.S.}
\newcommand{\gsu}{Georgia State University, Atlanta, GA 30303, U.S.}
\newcommand{\hiroshima}{Hiroshima University, Kagamiyama, Higashi-Hiroshima 739-8526, Japan}
\newcommand{\ihepprot}{IHEP Protvino, State Research Center of Russian Federation, Institute for High Energy Physics, Protvino, 142281, Russia}
\newcommand{\illuiuc}{University of Illinois at Urbana-Champaign, Urbana, IL 61801, U.S.}
\newcommand{\instpasczech}{Institute of Physics, Academy of Sciences of the Czech Republic, Na Slovance 2, 182 21 Prague 8, Czech Republic}
\newcommand{\isu}{Iowa State University, Ames, IA 50011, U.S.}
\newcommand{\jinrdubna}{Joint Institute for Nuclear Research, 141980 Dubna, Moscow Region, Russia}
\newcommand{\kek}{KEK, High Energy Accelerator Research Organization, Tsukuba, Ibaraki 305-0801, Japan}
\newcommand{\kfki}{KFKI Research Institute for Particle and Nuclear Physics of the Hungarian Academy of Sciences (MTA KFKI RMKI), H-1525 Budapest 114, POBox 49, Budapest, Hungary}
\newcommand{\korea}{Korea University, Seoul, 136-701, Korea}
\newcommand{\kurchatov}{Russian Research Center ``Kurchatov Institute", Moscow, Russia}
\newcommand{\kyoto}{Kyoto University, Kyoto 606-8502, Japan}
\newcommand{\labllr}{Laboratoire Leprince-Ringuet, Ecole Polytechnique, CNRS-IN2P3, Route de Saclay, F-91128, Palaiseau, France}
\newcommand{\lawllnl}{Lawrence Livermore National Laboratory, Livermore, CA 94550, U.S.}
\newcommand{\losalamos}{Los Alamos National Laboratory, Los Alamos, NM 87545, U.S.}
\newcommand{\lpc}{LPC, Universit{\'e} Blaise Pascal, CNRS-IN2P3, Clermont-Fd, 63177 Aubiere Cedex, France}
\newcommand{\lund}{Department of Physics, Lund University, Box 118, SE-221 00 Lund, Sweden}
\newcommand{\mass}{Department of Physics, University of Massachusetts, Amherst, MA 01003-9337, U.S. }
\newcommand{\muenster}{Institut f\"ur Kernphysik, University of Muenster, D-48149 Muenster, Germany}
\newcommand{\muhlenberg}{Muhlenberg College, Allentown, PA 18104-5586, U.S.}
\newcommand{\myongji}{Myongji University, Yongin, Kyonggido 449-728, Korea}
\newcommand{\nagasaki}{Nagasaki Institute of Applied Science, Nagasaki-shi, Nagasaki 851-0193, Japan}
\newcommand{\newmex}{University of New Mexico, Albuquerque, NM 87131, U.S. }
\newcommand{\nmsu}{New Mexico State University, Las Cruces, NM 88003, U.S.}
\newcommand{\ornl}{Oak Ridge National Laboratory, Oak Ridge, TN 37831, U.S.}
\newcommand{\orsay}{IPN-Orsay, Universite Paris Sud, CNRS-IN2P3, BP1, F-91406, Orsay, France}
\newcommand{\peking}{Peking University, Beijing, People's Republic of China}
\newcommand{\pnpi}{PNPI, Petersburg Nuclear Physics Institute, Gatchina, Leningrad region, 188300, Russia}
\newcommand{\riken}{RIKEN, The Institute of Physical and Chemical Research, Wako, Saitama 351-0198, Japan}
\newcommand{\rikjrbrc}{RIKEN BNL Research Center, Brookhaven National Laboratory, Upton, NY 11973-5000, U.S.}
\newcommand{\rikkyo}{Physics Department, Rikkyo University, 3-34-1 Nishi-Ikebukuro, Toshima, Tokyo 171-8501, Japan}
\newcommand{\saispbstu}{Saint Petersburg State Polytechnic University, St. Petersburg, Russia}
\newcommand{\saopaulo}{Universidade de S{\~a}o Paulo, Instituto de F\'{\i}sica, Caixa Postal 66318, S{\~a}o Paulo CEP05315-970, Brazil}
\newcommand{\seoulnat}{System Electronics Laboratory, Seoul National University, Seoul, Korea}
\newcommand{\stonybrkc}{Chemistry Department, Stony Brook University, Stony Brook, SUNY, NY 11794-3400, U.S.}
\newcommand{\stonycrkp}{Department of Physics and Astronomy, Stony Brook University, SUNY, Stony Brook, NY 11794, U.S.}
\newcommand{\subatech}{SUBATECH (Ecole des Mines de Nantes, CNRS-IN2P3, Universit{\'e} de Nantes) BP 20722 - 44307, Nantes, France}
\newcommand{\tenn}{University of Tennessee, Knoxville, TN 37996, U.S.}
\newcommand{\titech}{Department of Physics, Tokyo Institute of Technology, Oh-okayama, Meguro, Tokyo 152-8551, Japan}
\newcommand{\tsukuba}{Institute of Physics, University of Tsukuba, Tsukuba, Ibaraki 305, Japan}
\newcommand{\vandy}{Vanderbilt University, Nashville, TN 37235, U.S.}
\newcommand{\waseda}{Waseda University, Advanced Research Institute for Science and Engineering, 17 Kikui-cho, Shinjuku-ku, Tokyo 162-0044, Japan}
\newcommand{\weizmann}{Weizmann Institute, Rehovot 76100, Israel}
\newcommand{\yonsei}{Yonsei University, IPAP, Seoul 120-749, Korea}
\affiliation{\abilene}
\affiliation{\acadsin}
\affiliation{\banaras}
\affiliation{\barc}
\affiliation{\bnlchem}
\affiliation{\bnlcoll}
\affiliation{\bnlphys}
\affiliation{\caucr}
\affiliation{\charlesczech}
\affiliation{\ciae}
\affiliation{\cns}
\affiliation{\colorado}
\affiliation{\columbia}
\affiliation{\czechtech}
\affiliation{\dapnia}
\affiliation{\debrecen}
\affiliation{\elte}
\affiliation{\fit}
\affiliation{\fsu}
\affiliation{\gsu}
\affiliation{\hiroshima}
\affiliation{\ihepprot}
\affiliation{\illuiuc}
\affiliation{\instpasczech}
\affiliation{\isu}
\affiliation{\jinrdubna}
\affiliation{\kek}
\affiliation{\kfki}
\affiliation{\korea}
\affiliation{\kurchatov}
\affiliation{\kyoto}
\affiliation{\labllr}
\affiliation{\lawllnl}
\affiliation{\losalamos}
\affiliation{\lpc}
\affiliation{\lund}
\affiliation{\mass}
\affiliation{\muenster}
\affiliation{\muhlenberg}
\affiliation{\myongji}
\affiliation{\nagasaki}
\affiliation{\newmex}
\affiliation{\nmsu}
\affiliation{\ornl}
\affiliation{\orsay}
\affiliation{\peking}
\affiliation{\pnpi}
\affiliation{\riken}
\affiliation{\rikjrbrc}
\affiliation{\rikkyo}
\affiliation{\saispbstu}
\affiliation{\saopaulo}
\affiliation{\seoulnat}
\affiliation{\stonybrkc}
\affiliation{\stonycrkp}
\affiliation{\subatech}
\affiliation{\tenn}
\affiliation{\titech}
\affiliation{\tsukuba}
\affiliation{\vandy}
\affiliation{\waseda}
\affiliation{\weizmann}
\affiliation{\yonsei}
\author{A.~Adare}	\affiliation{\colorado}
\author{S.~Afanasiev}	\affiliation{\jinrdubna}
\author{C.~Aidala}	\affiliation{\columbia} \affiliation{\mass}
\author{N.N.~Ajitanand}	\affiliation{\stonybrkc}
\author{Y.~Akiba}	\affiliation{\riken} \affiliation{\rikjrbrc}
\author{H.~Al-Bataineh}	\affiliation{\nmsu}
\author{J.~Alexander}	\affiliation{\stonybrkc}
\author{K.~Aoki}	\affiliation{\kyoto} \affiliation{\riken}
\author{L.~Aphecetche}	\affiliation{\subatech}
\author{R.~Armendariz}	\affiliation{\nmsu}
% ...
\collaboration{PHENIX Collaboration} \noaffiliation

\date{\today}

\begin{abstract}
  Neutral pion transverse momentum ($p_\mathrm{T}$) spectra at
  midrapidity ($|y| \le 0.35$) were measured in Cu+Cu collisions at
  $\sqrt{s_{NN}}$ = 22.4, 62.4, and 200~GeV.  Relative to $\pi^0$
  yields in p+p collisions scaled by the number of inelastic
  nucleon-nucleon collisions ($N_\mathrm{coll}$) the $\pi^0$ yields
  for $p_\mathrm{T} \gtrsim 2$~GeV/$c$ in central Cu+Cu collisions are
  suppressed at $62.4$ and 200~GeV whereas an enhancement is observed
  at 22.4~GeV.  A comparison with a jet quenching model suggests that
  final state parton energy-loss dominates in central Cu+Cu collisions
  at 62.4~GeV and 200~GeV, while the enhancement at 22.4~GeV is
  consistent with nuclear modifications in the initial state alone.
\end{abstract}


\maketitle


The measurement of particle yields at high transverse momentum
($p_\mathrm{T} \gtrsim 2$~GeV/$c$) has played a key role in
characterizing the medium created in nucleus-nucleus collisions at the
Relativistic Heavy Ion Collider (RHIC)
\cite{Adcox:2004mh,Adare:ppg079}. Hadrons produced at sufficiently
high $p_\mathrm{T}$ result from the interaction of quarks and gluons
with high momentum transfer (``hard scattering'') which can be
described by perturbative quantum-chromodynamics (pQCD). These hadrons
are produced as particle jets in the fragmentation of the scattered
partons. A scattered parton propagating through a quark-gluon plasma,
a thermalized medium in which quarks and gluons are not confined in
hadrons, loses energy (``jet-quenching'') resulting in hadron yields
at high $p_\mathrm{T}$ being suppressed \cite{Gyulassy:1990ye}.  Such
a suppression was indeed observed in central Au+Au collisions at
$\sqrt{s_{NN}} = 130$ and 200~GeV at RHIC, providing evidence
for large color-charge densities in these systems
\cite{Adcox:2001jp,Adler:2003qi,Adams:2003kv}.

Characteristic properties of the suppression of hadrons at
high-$p_\mathrm{T}$, {\it e.g.}, the dependence on $p_\mathrm{T}$ and
centrality, were studied in detail in Au+Au collisions at
$\sqrt{s_{NN}} = 200$~GeV \cite{Adler:2003qi}.  However, the
energy dependence of hadron production in A+A collisions as predicted
by jet quenching models \cite{Wang:1998ww,Vitev:2002pf,Vitev:2004gn}
is not well constrained by measurements. Work in this direction was
presented in \cite{Alver:2005nb,Abelev:2007ra,Aggarwal:2007gw}.  To
study the energy dependence of jet-quenching it is desirable to
measure identified particles in the same colliding system over a large
$\sqrt{s_{NN}}$ range and to compare to p+p reference data
measured in the same experimental setup.  Identified particles provide
an advantage over unidentified hadrons in that the interpretation is
not complicated by the different contributions from baryons and
mesons.  The study of Cu+Cu collisions is particularly useful because
hadron suppression in Au+Au collisions is observed for rather
peripheral collisions with a number of participating nucleons of
$N_\mathrm{part} \sim 50-100$ \cite{Adler:2003qi}.  This
$N_\mathrm{part}$ range can be studied with reduced uncertainties in
$N_\mathrm{coll}$ with the smaller $^{63}$Cu nucleus.

A critical parameter in jet quenching models is the initial
color-charge density of the medium. By studying Cu+Cu collisions in
the range $\sqrt{s_{NN}} \sim 20-200$~GeV this parameter can be
varied with essentially no change in transverse size and shape of the
reaction zone.  Moreover, the enhancement of hadron yields due to
multiple soft scattering of the incoming partons (``nuclear
$k_\mathrm{T}$'' or ``Cronin enhancement'') is expected to increase
towards smaller $\sqrt{s_{NN}}$ \cite{Vitev:2002pf}, thus the
interplay between this enhancement and the suppression due to parton
energy-loss can be studied.

In this letter we present invariant $\pi^0$ yields for Cu+Cu
collisions at $\sqrt{s_{NN}} = 22.4, 62.4$, and 200~GeV.
Reference data for p+p collisions at $\sqrt{s} = 62.4$~GeV and 200~GeV
were taken with the same experiment \cite{Adare:2007zz,Adare:2007dg}.
At $\sqrt{s_{NN}} = 22.4$~GeV p+p reference data were obtained
from a parameterization of the world's data on $\pi^0$ production
\cite{Arleo:2008}.


Neutral pions were measured via their $\pi^0 \rightarrow \gamma\gamma$
decay branch with the electromagnetic calorimeter (EMCal) of the
PHENIX experiment \cite{Aphecetche:2003zr}. The EMCal comprises two
calorimeter types: 6 sectors of a lead scintillator sampling
calorimeter (PbSc) and 2 sectors of a lead glass Cherenkov calorimeter
(PbGl).  Each sector is located $\sim 5$\,m from the beamline and
subtends $|\eta| < 0.35$ in pseudorapidity and $\Delta \varphi =
22.5^\circ$ in azimuth.  Owing to the PbSc (PbGl) granularity of
$\Delta \eta \times \Delta \varphi = 0.011 \times 0.011$ ($0.008
\times 0.008$) the probability that the two photon showers from a
$\pi^0$ decay result in partially overlapping clusters is negligible
up to a $\pi^0$ $p_\mathrm{T}$ of 12~GeV/$c$ (15~GeV/$c$). The
energy calibration of the EMCal was corroborated by the position of
the $\pi^0$ invariant mass peak, the energy deposited by minimum
ionizing charged particles traversing the EMCal (PbSc), and the
correlation between the measured momenta of electron and positron
tracks identified by the ring-imaging Cherenkov detector and the
associated energy deposited in the EMCal.  These studies showed that
the accuracy of the energy measurement was better than 1.5\%.


The total number of analyzed Cu+Cu events for the three energies is
shown in Table~\ref{tab:datasets}. The minimum bias (MB) trigger for
all reaction systems was provided by Beam-Beam-Counters (BBC's)
located at $3.0 \lesssim |\eta| \lesssim 3.9$. The reaction vertex
along the beam axis, determined from the arrival time differences in
the BBC's, was required to be in the range $|z| \le 30$~cm. An
additional hardware trigger (ERT) on high-$p_\mathrm{T}$
photons/electrons was employed in Cu+Cu at $\sqrt{s_{NN}} =
200$~GeV. This trigger was based on the analog energy signal measured
in overlapping $4 \times 4$ towers of the EMCal in coincidence with
the MB trigger condition.  The ERT reached a efficiency plateau
for photon energies $E \gtrsim 4$~GeV.

\begin{table}[t]
  \caption{\label{tab:datasets} Data sets presented in this paper with
    the number of analyzed events. For the ERT triggered data the number
    of equivalent minimum bias events is given. }
\begin{ruledtabular}
\begin{tabular}{ccccc}
  system & 
  $\sqrt{s_{NN}}$ & 
  $\varepsilon_\mathrm{trig}$ & 
  $N_\mathrm{evt}^\mathrm{MB}$ & 
  $N_\mathrm{evt}^\mathrm{ERT}$ ($N_\mathrm{evt}^\mathrm{sampled}$) \\ 
  \hline
  Cu+Cu & 22.4 GeV & $75-90\%$        & $5.8 \cdot 10^{6}$ &  --- \\
  Cu+Cu & 62.4 GeV & $(88 \pm 4) \%$ & $192 \cdot 10^{6}$ &  --- \\
  Cu+Cu & 200 GeV  & $(94 \pm 2) \%$ & $794 \cdot 10^{6}$ &  $15.5 \cdot 10^{6}$ ($4720 \cdot10^{6}$)\\
\end{tabular}
\end{ruledtabular}
\end{table}


The centrality selection in Cu+Cu at $\sqrt{s_{NN}} = 200$~GeV
and $\sqrt{s_{NN}} = 62$~GeV was based on the charge signal of
the BBC's which is proportional to the charged particle multiplicity
in the respective pseudorapidity range. The BBC trigger efficiency
($\varepsilon_\mathrm{trig}$) for these systems was determined with
the aid of the HIJING event generator and a full GEANT simulation of
the BBC response (see Table~\ref{tab:datasets}).  At
$\sqrt{s_{NN}} = 22.4$~GeV centrality classes were defined
based on the charged particle multiplicity measured with the pad
chamber (PC1) detector ($|\eta| < 0.35$).  The measured PC1
multiplicity distribution was accurately reproduced in a Glauber Monte
Carlo calculation \cite{Miller:2007ri} and centrality classes were
determined by identical cuts on the measured and simulated PC1
multiplicities. The estimated BBC trigger efficiency given in
Table~\ref{tab:datasets} results from a comparison of the simulated
and the measured PC1 multiplicity distributions.  The results of the
Glauber calculation \cite{Miller:2007ri} for Cu+Cu collisions at 22.4,
62.4, and 200~GeV using inelastic cross sections of 32.3, 35.6, and 42
mb, respectively, are given in Table~\ref{tab:npartncoll}.
\begin{table}[t]
  \caption{\label{tab:npartncoll}
    Glauber Monte Carlo calculations for Cu+Cu collisions at 
    22.4, 62.4, and 200 GeV.
    The $N_\mathrm{coll}$ systematic uncertainty at 62.4 and 200 GeV 
    is $\sim 12\%$, almost independent of $N_\mathrm{coll}$. 
    At 22.4 GeV the relative uncertainty of $N_\mathrm{coll}$
    can be parameterized as $0.094 + 0.173 e^{-0.0272 N_\mathrm{coll}}$.
  }
\begin{ruledtabular}
\begin{tabular}{lcccccc}
  & 
  \multicolumn{2}{c}{22.4\,GeV} & 
  \multicolumn{2}{c}{62.4\,GeV} & 
  \multicolumn{2}{c}{200\,GeV} \\ 
  &  
  $\langle N_\mathrm{part} \rangle$ & 
  $\langle N_\mathrm{coll} \rangle$ & 
  $\langle N_\mathrm{part} \rangle$ & 
  $\langle N_\mathrm{coll} \rangle$ & 
  $\langle N_\mathrm{part} \rangle$ & 
  $\langle N_\mathrm{coll} \rangle$ \\  
  \hline
   0-10\,\% & 92.2 & 140.7 & 93.3 & 152.3 & 98.2 & 182.7 \\
  10-20\,\% & 67.8 &  93.3 & 71.1 & 105.5 & 73.6 & 121.1 \\
  20-30\,\% & 48.3 &  59.7 & 51.3 &  67.8 & 53.0 &  76.1 \\
  30-40\,\% & 34.1 &  38.0 & 36.2 &  42.6 & 37.3 &  47.1 \\
  40-50\,\% & 23.1 &  22.9 & 24.9 &  26.2 & 25.4 &  28.1 \\
  50-60\,\% & 15.5 &  13.9 & 16.1 &  15.0 & 16.7 &  16.2 \\
  60-70\,\% & ---  & ---   & ---  &  ---  & 10.4 &   9.0 \\
  70-80\,\% & ---  & ---   & ---  &  ---  &  6.4 &   4.9 \\
  80-94\,\% & ---  & ---   & ---  &  ---  &  3.6 &   2.4 \\
  60-88\,\% & ---  & ---   & 7.0  & 5.5   & ---  &  ---  \\
\end{tabular}
\end{ruledtabular}
\end{table}


Neutral pions yields were measured on a statistical basis by
calculating the invariant mass of all photon pairs in a given event
and counting those within the $\pi^0$ mass range. The background of
combinatorial pairs was calculated by pairing photon hits from
different events. Only photon pairs with an energy asymmetry
$|E_1-E_2|/(E_1+E_2) < 0.7$ were accepted. The raw $\pi^0$ yields were
corrected for the geometrical acceptance and reconstruction
efficiency.  The latter takes into account the loss of $\pi^0$'s due
to photon identification cuts, the energy asymmetry cut, inactive
detector areas, and photon conversions.  Moreover, it corrects the
distortion of the $\pi^0$ spectrum which results from the finite
energy resolution in conjunction with the steeply falling spectra and
shower overlap effects.  For Cu+Cu at $\sqrt{s_{NN}} = 200$~GeV
the transition between the minimum bias and the ERT sample occurs at
$p_\mathrm{T} = 8$~GeV/$c$. The final spectra were calculated as the
weighted average of the PbSc and PbGl results, which agree well within
the uncertainties.

\begin{figure*}
\includegraphics[width=1.0\linewidth]{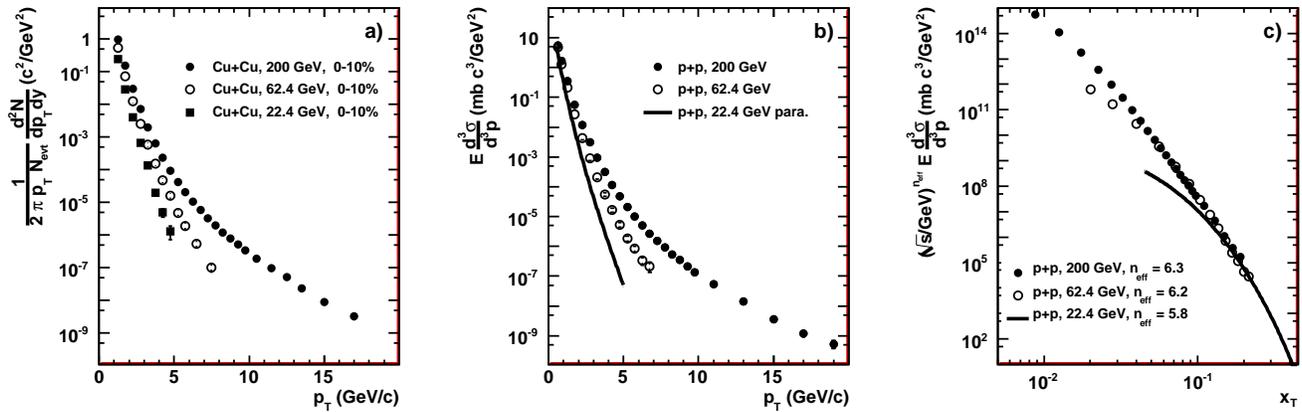}  
\caption{\label{fig:spectra} Invariant $\pi^0$ yields in central Cu+Cu
  collisions (a) and invariant $\pi^0$ cross sections in p+p
  collisions (b) at $\sqrt{s_{NN}} = 22.4, 62.4, 200$~GeV
  \cite{Adare:2007zz,Adare:2007dg}. The error bars represent the
  quadratic sum of the statistical and total systematic uncertainties.
  Plotted as a function of $x_\mathrm{T} = 2 p_\mathrm{T}/\sqrt{s}$
  (c) the p+p data exhibit an approximate $x_\mathrm{T}$ scaling.} 
\end{figure*}


The main systematic uncertainties of the $\pi^0$ spectra result from
the $\pi^0$ peak extraction, the reconstruction efficiency, and the
EMCal energy calibration. For $p_\mathrm{T} \gtrsim 2$~GeV/$c$ the
peak extraction uncertainty is $\sim 4\,\%$ for all systems,
approximately independent of $p_\mathrm{T}$. The uncertainty in the
reconstruction efficiency was estimated to be $\sim 15\,\%$ for the
three Cu+Cu analyses. It includes uncertainties of the photon
identification cuts, the energy resolution, and the modeling of shower
overlap effects. The uncertainty in the EMCal energy scale of
$1.5\,\%$ translates into an uncertainty in the yields that increases
from $\sim 8\,\%$ at $p_\mathrm{T} = 3$~GeV/$c$ to $15\,\%$ at
$p_\mathrm{T} = 6$~GeV/$c$. The high-$p_\mathrm{T}$ part of the
spectra in Cu+Cu at 200~GeV measured with the ERT trigger is subject
to an additional uncertainty of $10\,\%$ related to the ERT trigger
efficiency and normalization.


PHENIX has not yet acquired a p+p data set at $\sqrt{s} = 22.4$~GeV.
Therefore world data on charged and neutral pion production in the
range $21.7 \le \sqrt{s} \le 23.8$~GeV were scaled to $\sqrt{s} =
22.4$~GeV and fit in the range $0 \lesssim p_\mathrm{T} \lesssim
7$~GeV/$c$ with $ E \, \mathrm{d^3\sigma}/\mathrm{d}p^3 = A (e^{a
  p_\mathrm{T}+b})^n (\sqrt{s}/2 - p_\mathrm{T})^m$ where $A = 1.22
\cdot 10^{-17}$~mb\,GeV$^{-2}c^3$, $a = 0.053$~GeV$^{-1}c$, $b =
-0.884$, $n = -15.25$, and $m = 4.653$ \cite{Arleo:2008}.  The scaling
correction was determined with a next-to-leading-order QCD
calculation. The scaling correction was largest for $\sqrt{s} =
23.8$~GeV and reduced these spectra by $\sim 30\%$ \cite{Arleo:2008}.
The parameterization is consistent within $\pm 25\,\%$ with the
existing $\pi^0$ and $\pi^{\pm}$ measurements without discernible
$p_\mathrm{T}$-dependent systematic deviations.


The $\pi^0$ $p_\mathrm{T}$ spectra for p+p and central Cu+Cu
collisions ($0-10$\% of $\sigma_\mathrm{inel}^\mathrm{Cu+Cu}$) at
$\sqrt{s_{NN}} = 22.4, 62.4$~\cite{Adare:2007zz}, and 200~GeV
\cite{Adams:2003kv} are shown in Fig.~\ref{fig:spectra}a and
\ref{fig:spectra}b. At sufficiently high $p_\mathrm{T}$ where pion
production in p+p collisions is dominated by fragmentation of jets,
QCD predicts a scaling law
$\sqrt{s}^{n_\mathrm{eff}(x_\mathrm{T},\sqrt{s})} \, E
\mathrm{d}^3\sigma/\mathrm{d}p^3 = G(x_\mathrm{T})$ with a universal
function $G(x_\mathrm{T})$ where $x_\mathrm{T} = 2
p_\mathrm{T}/\sqrt{s}$ \cite{Cahalan:1974tp}.  Fig.~\ref{fig:spectra}c
shows that such a scaling in $x_\mathrm{T}$ is indeed observed for p+p
collisions at 22.4, 62.4, and 200~GeV, consistent with previous
observations \cite{Adler:2003au}.  The $x_\mathrm{T}$ values at which
the universal curve $G(x_\mathrm{T})$ is reached indicate that
particle production is dominated by hard processes for $p_\mathrm{T}
\gtrsim 2$~GeV/$c$ for the three considered energies.

\begin{figure}[t]
\includegraphics[width=1.0\linewidth]{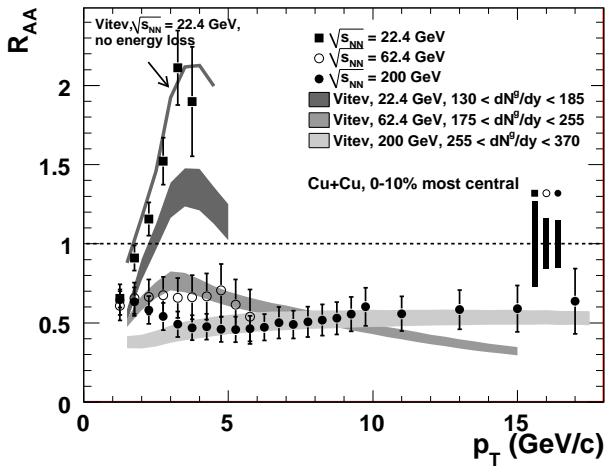}
\caption{
\label{fig:raa_pt_0-10}
Measured $\pi^0$ $R_{AA}$ as a function of $p_\mathrm{T}$ for the
$0-10\%$ most central Cu+Cu collisions at $\sqrt{s_{NN}} =
22.4, 62.4, 200$~GeV in comparison to a jet quenching calculation
\cite{Vitev:2005he, Vitev:privcom}.  The error bars in this figure
(and in Fig.~\ref{fig:raa_vs_npart}) represent the quadratic sum of
the statistical uncertainties and the point-to-point uncorrelated and
correlated systematic uncertainties.  The boxes around unity indicate
uncertainties related to $\langle N_\mathrm{coll} \rangle$ and
absolute normalization.  The bands for the theory calculation
correspond to the assumed range of the initial gluon density
$\mathrm{d}N^\mathrm{g}/\mathrm{d}y$. The thin solid line is a
calculation without parton energy-loss for central Cu+Cu at
$\sqrt{s_{NN}} = 22.4$~GeV.}
\end{figure}

Nuclear effects on high-$p_\mathrm{T}$ $\pi^0$ production can be quantified 
with the nuclear modification factor 
\begin{equation}
  R_\mathrm{AA}(p_\mathrm{T})\,=\,
  \frac{(1/N^\mathrm{evt}_\mathrm{AA})\,\mathrm{d}^2N_\mathrm{AA}/\mathrm{d}p_\mathrm{T} dy}
  {\langle T_\mathrm{AB} \rangle 
    \,\times\, \mathrm{d}^2\sigma_\mathrm{pp}/\mathrm{d}p_\mathrm{T} dy} 
\label{eq:R_AA}
\end{equation}
where $\langle T_\mathrm{AB} \rangle = \langle
N_\mathrm{coll}\rangle/\sigma_\mathrm{pp}^\mathrm{inel}$.  In the
absence of nuclear effects $R_\mathrm{AA} = 1$ for $p_\mathrm{T}
\gtrsim 2$~GeV/$c$ where pions result from hard scattering processes.
$R_\mathrm{AA}(p_\mathrm{T})$ for the $0-10\%$ most central Cu+Cu
collisions at 22.4, 62.4, and 200 GeV is shown in
Fig.~\ref{fig:raa_pt_0-10}. The suppression at 62.4 GeV
($R_\mathrm{AA} \approx 0.6$ for $p_\mathrm{T} \gtrsim 3$~GeV/$c$) and
200 GeV ($R_\mathrm{AA} \approx 0.5 - 0.6$ for $p_\mathrm{T} \gtrsim
3$~GeV/$c$) is consistent with expectations from parton energy-loss.
The $R_\mathrm{AA} > 1$ in Cu+Cu at 22.4~GeV is similar to the
enhancement by a factor $\sim 1.5$ (at $p_\mathrm{T} \approx
3$~GeV/$c$) observed in p+W relative to p+Be collisions at
$\sqrt{s_{NN}} = 19.4$~GeV and 23.8~GeV
\cite{Antreasyan:1978cw}.  For a similar number of participants the
$R_\mathrm{AA}$ in Cu+Cu at 22.4~GeV agrees with the $R_\mathrm{AA}$
in Pb+Pb collisions at 17.3~GeV \cite{Aggarwal:2007gw}.


For $p_\mathrm{T} \gtrsim 3$~GeV/$c$ the measured nuclear modification
factors at 62.4, and 200~GeV are consistent with a numerically
evaluated parton energy-loss model described in
\cite{Vitev:2005he,Vitev:privcom} as indicated by the comparison in
Fig.~\ref{fig:raa_pt_0-10}. This calculation takes into account
shadowing from coherent final state interactions in nuclei
\cite{Qiu:2004da}, Cronin enhancement \cite{Vitev:2003xu}, initial
state parton energy-loss in cold nuclear matter \cite{Vitev:2007ve},
and final state parton energy-loss in dense partonic matter
\cite{Vitev:2004gn, Vitev:2005he, Vitev:privcom}.  The Cronin
enhancement measured in p+A collisions is described well by this model
\cite{Vitev:2003xu}.  The initial gluon rapidity density
$\mathrm{d}N^\mathrm{g}/\mathrm{d}y$ which characterizes the medium
was not fit to the $R_\mathrm{AA}$ values, but instead was constrained
by measured charged-particle multiplicities and the assumption of
parton-hadron duality ($\mathrm{d}N^\mathrm{g}/\mathrm{d}y = \kappa \,
\mathrm{d}\eta/\mathrm{d}y \, \mathrm{d}N_\mathrm{ch}/\mathrm{d}\eta $
with $\kappa = 3/2 \pm 30\%$ and $\mathrm{d}\eta/\mathrm{d}y \equiv
1.2$ at all energies) \cite{Vitev:2005he, Vitev:privcom}.  The average
fractional energy losses $\Delta E/E$ for a quark (gluon) with $E =
6$~GeV corresponding to the $\mathrm{d}N^\mathrm{g}/\mathrm{d}y$
ranges in Fig.~\ref{fig:raa_pt_0-10} are $0.13-0.19$ ($0.29-0.42$),
$0.16-0.20$ ($0.35-0.44$), $0.20-0.28$ ($0.44-0.63$) in central Cu+Cu
collisions at 22.4, 62.4, and 200 GeV, respectively
\cite{Vitev:privcom}.  For Cu+Cu at $\sqrt{s_{NN}} = 22.4$~GeV
the calculation is also shown without final state parton energy-loss.
The measurement is consistent with this calculation but does not rule
out a scenario with parton energy-loss.

\begin{figure}
\includegraphics[width=1.0\linewidth]{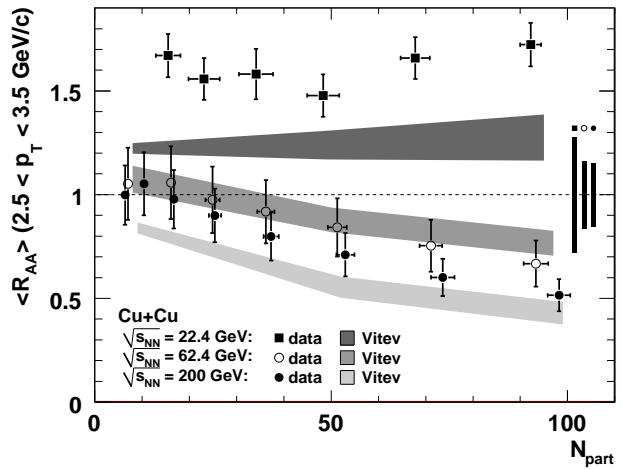} 
\caption{\label{fig:raa_vs_npart}
The average $R_{AA}$ in the interval $2.5 < p_\mathrm{T} <
3.5$~GeV/$c$ as a function of centrality for Cu+Cu collisions at
$\sqrt{s_{NN}} = 22.4, 62.4$, and 200~GeV. The shaded bands
represent jet quenching calculations at three discrete centralities
($N_\mathrm{part} \sim 10, 50, 100$) \cite{Vitev:2005he,
  Vitev:privcom}. The boxes around unity represent the normalization
and $\langle N_\mathrm{coll}\rangle$ uncertainties for a typical
$N_\mathrm{coll}$ uncertainty of 12\,\%.}
\end{figure}

Fig.~\ref{fig:raa_vs_npart} shows that the $\pi^0$ suppression in the
range $2.5 < p_\mathrm{T} < 3.5$~GeV/$c$ increases towards more
central Cu+Cu collisions for $\sqrt{s_{NN}} = 62.4$, 200~GeV.
On the other hand, $R_\mathrm{AA}$ at $\sqrt{s_{NN}} =
22.4$~GeV remains approximately constant as a function of
$N_\mathrm{part}$, suggesting either that the Cronin enhancement
depends only weakly on centrality or that in this energy range parton
energy-loss is offset by the larger effect of Cronin enhancement over
a broad range of centrality. It appears from these data that in Cu+Cu
collisions between $\sqrt{s_{NN}} = 22.4$ and 62.4~GeV parton
energy-loss will start to prevail over the Cronin enhancement,
resulting in a net suppression.

%
In summary, for the first time $\pi^0$ $p_\mathrm{T}$ spectra for the
same nuclear colliding system (Cu+Cu) were measured in the same
experimental setup over a wide range of energies
($\sqrt{s_{NN}} = 22.4, 62.4$, and 200~GeV).  Nuclear effects
were studied using measured p+p $\pi^0$ reference spectra from PHENIX
at 62.4 and 200~GeV, and a parameterization of world data at 22.4~GeV.
High-$p_\mathrm{T}$ $\pi^0$ yields in central Cu+Cu collisions at
62.4~GeV and 200~GeV are suppressed, suggesting that parton
energy-loss is a significant effect in these systems.  At 22.4~GeV
$\pi^0$ yields for $p_\mathrm{T} \gtrsim 2$~GeV/$c$ are not
suppressed.  The $R_\mathrm{AA}$ measured in central Cu+Cu at 22.4 GeV
is consistent with Cronin enhancement alone but does not rule out
parton energy-loss effects.  The measurements of high-$p_\mathrm{T}$
$\pi^0$ production over a factor $\sim 10$ in center-of-mass energy
presented in this letter provide a unique constraint for jet-quenching
models and demonstrate that parton energy-loss starts to prevail over
the Cronin enhancement between $\sqrt{s_{NN}} = 22.4$ and
62.4~GeV.

%

We thank the staff of the Collider-Accelerator and 
Physics Departments at BNL for their vital contributions.  
We thank Ivan Vitev for
providing the jet quenching calculations. 
We acknowledge support from 
the Office of Nuclear Physics in DOE Office of Science and NSF (U.S.A.), 
MEXT and JSPS (Japan), 
CNPq and FAPESP (Brazil), 
NSFC (China), 
MSMT (Czech Republic),
IN2P3/CNRS, and CEA (France), 
BMBF, DAAD, and AvH (Germany), 
OTKA (Hungary), 
DAE (India), 
ISF (Israel), 
KRF and KOSEF (Korea), 
MES, RAS, and FAAE (Russia),
VR and KAW (Sweden), 
U.S. CRDF for the FSU, 
US-Hungary Fulbright, 
and US-Israel BSF.


\end{document}